\documentclass[%
 reprint,
 amsmath,amssymb,
 aps,
]{revtex4-2}

\usepackage[utf8]{inputenc}%
\usepackage{amssymb}
\usepackage{siunitx}
\usepackage{amsmath}
\usepackage{float}
\usepackage{hyperref}
\hypersetup{hidelinks,colorlinks=true,linkcolor=blue,citecolor=blue,urlcolor=black}
\usepackage{graphicx}
\usepackage{dcolumn}
\usepackage{bm}

\usepackage{url}
\begin{document}

\preprint{APS/123-QED}

\title{Magnetically and electrically controllable valley splittings in MXene monolayers  }

\author{Huiqian Wang}
\author{Li Liang}
\author{Xiaohui Wang}
\author{Xiaoyu Wang}
\author{Xiao Li}%
 \email{lixiao@njnu.edu.cn}
\affiliation{%
 Center for Quantum Transport and Thermal Energy Science,\\
 School of Physics and Technology, Nanjing Normal University, Nanjing 210023, China 
}%




\date{\today}

\begin{abstract}
The modulation of the valley structure in two-dimensional valley materials is vital in the field of valleytronics. The multiferroicity provides possibility for multiple modulations of the valley, including the magnetic and electric means. Based on the first-principle calculations, we study the valley properties and associated manipulations of multiferroic Co$_2$CF$_2$ monolayers with different stacking patterns. Our calculations show that the Co$_2$CF$_2$ monolayer in the H$^{\prime}$ phase is a ferrovalley material, with sizable valley splittings. By rotating the magnetization direction, the valley splittings can be tuned for both the magnitude and sign. The electric field, driving the reversal of the electric polarization, can also change the magnitude of the valley splittings. Besides, a metastable T$^{\prime}$ phase exhibits valley splittings as well, of which the magnitude and sign can be simultaneously controlled by applied magnetic and electric fields. These findings offer a practical way for realizing highly tunable valleys by multiferroic couplings. 
\end{abstract}

\maketitle


\section{INTRODUCTION}
The electronic valley degree of freedom in two-dimensional materials has attracted immense attentions in recent years \cite{1,2,3,4,valley}. The key to exploit the valley degree of freedom relies on effectively tuning the valley degeneracy. The valley degeneracy is found to be modulated by the external magnetic field or the proximity-induced Zeeman effect in monolayer transition metal dichalcogenides, such as MoSe$_2$ and WSe$_2$ \cite{5,6,7,8,9}. As for two-dimensional materials with the valley-layer locking, e.g. the MoS$_2$ bilayer and the TiSiCO monolayer, applied electric field can also be used to tune the valley structures \cite{10,11,12}. There is an attractive question, i.e. whether the valley structure in a two-dimensional material can be modulated by both the magnetic and electric means. If so, the valley degree of freedom will be highly tunable and have enormous application uses in valleytronics. However, simultaneous magnetic and electric controls of the valley structure have few examples, and they are worth further exploration.    

Layered transition metal carbides and nitrides (MXenes) are a large branch of two-dimensional materials \cite{13,14,15}, and they exhibit a variety of exceptional properities, such as high mechanical strength, superconductivity and topolgical insulating \cite{16,17,18}. In particually, the multiferroicity in MXenes has been investigated recently. For example, the Co$_2$CF$_2$, Hf$_2$VC$_2$F$_2$ and Ti$_3$C$_2$T$_x$ monolayers are identified to be multiferroic materials, with both ferroelectric and ferromagnetic orders \cite{19,20,21}. Furthermore, if the valley structure can be found in a multiferroic MXene monolayer, the monolayer will become more compelling. It is likely to realize more couplings between multiferroic orders and correspondingly enable the magnetic and electric controls of the valley degree of freedom. 

In this work, taking the H$^{\prime}$-Co$_2$CF$_2$ monolayer for example, we study the valley properties and associated manipulations in multiferroic MXenes by first-principles density functional theory calculations. Our calculations show that there are two inequivalent $K_\pm$ valleys in the electronic band structure of the H$^{\prime}$-Co$_2$CF$_2$ monolayer, with considerable valley splittings for both the valence and conduction bands.
As a result, the monolayer is also a ferrovalley material besides ferroelectricity and ferromagnetism. Subsequently, the multiferroic couplings between ferroelectricity, ferromagnetism and ferrovalley are found. The valley splittings can be tuned by the magnetization rotation and applied electric field. Moreover, we also discuss that the electric control of the valleys in the T$^{\prime}$ phase of the Co$_2$CF$_2$ monolayer and another MXenes.   
The highly tunable valley splittings in multiferroic MXenes provides a practical avenue for designing advanced valleytronic and spintronic devices based on the couplings between multiferroic orders.                 

\section{METHODS}
We perform density functional theory calculations to study the atomic and electronic structures of the Co$_2$CF$_2$ monolayer. The calculations are implemented in the Vienna Ab initio Simulation Package \cite{22,23}, with the projector-augmented wave potentials and Perdew-Burke-Ernzerhof functional \cite{24,25}. A plane-wave energy cutoff of 600 eV and a Monkhorst-Pack \textbf{k}-point mesh of 13 × 13 × 1 are adopted. A vacuum slab of about 30 \AA \ is inserted to model the two-dimensional system. The convergence criterion for the total energy of electron iterations is set to 10$^{-6}$ eV. The atomic structure is fully relaxed until each interactomic force is less than 0.01 eV/\AA. An effective Hubbard correction, $U_\text{eff}$=2 eV, is considered to better describe the electron-electron interaction \cite{U}. For the asymmetric monolayer, the dipole correction is also included in our calculations. The electric polarizations of ferroelectric phases are calculated by using the Berry phase method \cite{26}. The energy barrier during the ferroelectric switching of the Co$_2$CF$_2$ monolayer is obtained by the nudging elastic band method \cite{27}. The Berry curvature is calculated using the WANNIER90 package \cite{28}. 

For the valley splitting focused on in our work, we define it as the energy difference between the $K_+$ and $K_-$ valleys. It reads,
\begin{equation}
{\Delta}^{v/c}=E_{K_+}^{v/c}-E_{K_-}^{v/c}
\end{equation}
where the superscripts $v$ and $c$ denote the valence and conduction bands, respectively. $E_{K\pm}^{v/c}$ is the energy extremum of the band-edge state at a given valley for a certain band.

\section{RESULTS}
\subsection{\label{sec:level1}Atomic structures and magnetic properties}
\begin{figure}[htbp]
\centering
\includegraphics[width=85 mm]{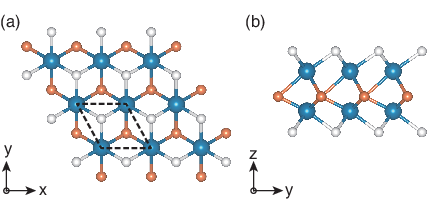}
\caption{\label{FIG. 1}  
Atomic structures of the H$^{\prime}$-Co$_2$CF$_2$ monolayer. (a) The top view and (b) the side view. Blue, orange and white  balls stand for Co, C and F atoms, respectively. The unit cell is bounded by dashed lines.}
\end{figure}

In our calculations, we first consider the most stable H$^{\prime}$ phase of the Co$_2$CF$_2$ monolayer. Fig.~\ref{FIG. 1} shows the atomic structure of the relaxed H$^{\prime}$-Co$_2$CF$_2$ monolayer. The monolayer has a two-dimensional hexagonal lattice, and consists of five atomic layers stacked as F-Co-C-Co-F. The middle C atomic layer is closer to one Co atomic layer than to the other Co atomic layer. Therefore, the H$^{\prime}$-Co$_2$CF$_2$ monolayer has no horizontal mirror symmetry. The in-plane lattice constant of its hexagonal lattice is computed to 2.90 {\AA} and the thickness of the monolayer is 4.50 \AA. The vertical distances between the middle C atomic layer and the two Co layers have a difference of 0.31 \AA. These calculation results agree with the previous study \cite{19}. Besides the H$^{\prime}$ phase, there is also a H phase with a mirror symmetry, in which the C atomic layer is exactly in the middle of the monolayer. According to the total energy calculations, it is found that the H$^{\prime}$ phase has a lower energy than the H phase.

Given that the H$^{\prime}$-Co$_2$CF$_2$ monolayer has a ferromagnetic order, we have also calculated its magnetic moment. It is found that the Co atom farther away from the middle C atomic layer has a magnetic moment of about 2 $\mu_\text{B}$, while the magnetic moment of the Co atom on the other side of the middle C atomic layer is vanishing. In contrast, each Co atom in the mirror-symmetric H phase has a magnetic moment of about 1 $\mu_\text{B}$. The calculated magnetic moments above are also consistent with the previous results \cite{19}.    

\subsection{\label{sec:bandstructure}Electronic band structures}
\begin{figure}[htbp]
\centering
\includegraphics[width=85 mm]{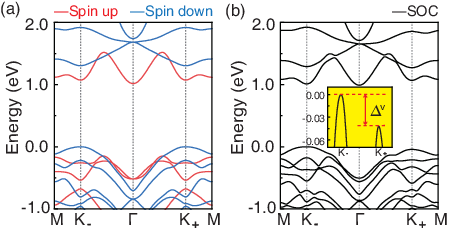} 
\caption{\label{FIG. 2}
Electronic band structures of the H$^{\prime}$-Co$_2$CF$_2$ monolayer. \label{(a)} The spin-orbit coupling is not included. \label{(b)} The spin-orbit coupling is included. In (a), red and blue lines represent spin-up and spin-down bands, respectively. The inset of (b) zooms in the valence band at $K_\pm$ valleys, with $\Delta^v$ denoting its valley splitting.}
\end{figure}

In the followings, we focus on electronic band structures and associated valley properties of the H$^{\prime}$-Co$_2$CF$_2$ monolayer, which are shown in Fig.~\ref{FIG. 2}. For the nonrelativistic band structure in Fig.~\ref{FIG. 2} (a), it is seen that the bands have spin splittings due to the magnetic exchange interaction. The band-edge states of the valence and conduction bands correspond to the spin-down and spin-up ones, respectively. The opposite spins in the two bands indicates that the monolayer is a bipolar magnetic semiconductor \cite{29}. On the other hand, it is seen that the degenerate valence band maxima (VBM) simultaneously appears at the $K_+$ and $K_-$ points of the Brillouin zone. At these two points, there are also two degenerate local minima of the conduction band, while the conduction band minimum (CBM) is localed at $\Gamma$ point. Therefore, the $K_+$ and $K_-$ are two degenerate, inequivalent valleys. Besides, given that the VBM and CBM are located at $K_\pm$ and $\Gamma$, respectively, the monolayer is an indirect-band-gap semiconductor, with a band gap of \SI{1.02}{eV}. The direct band gap has a magnitude of \SI{1.07}{eV} at $K_\pm$. Moreover, the second conduction band has spin-down global minima, which are also located at $K_\pm$ valleys. The direct band gap between the valence band and the second conduction band is \SI{1.30}{eV} at $K_\pm$ valleys.

Fig.~\ref{FIG. 2} (b) shows the band structure of the H$^{\prime}$-Co$_2$CF$_2$ monolayer with taking into account the spin-orbit coupling (SOC). It is seen that there are valley degeneracy splittings between the $K_+$ and $K_-$ valleys for both valence and conduction bands. That is, for a certain band, the band energies at $K_\pm$ valleys become different. The valley splittings arise from the cooperative roles of the SOC and the exchange interaction \cite{9}. According to the definition of the valley splitting in the method section, the valley splitting of the valence band is calculated to be \SI{-42.0} {meV}, while the first and second conduction bands have valley splittings of \SI{57.0} {meV} and \SI{-18.0} {meV}, respectively. Therefore, the H$^{\prime}$-Co$_2$CF$_2$ monolayer is indeed a ferrovalley material with considerable valley splittings \cite{30}. On the other hand, the magnitudes of the band gaps are changed after considering the SOC. The direct band gaps become 1.11 eV at $K_+$ valley and 1.01 eV at $K_-$ valley. Given that the VBM and CBM are localed at $K_-$ and $\Gamma$ with the SOC considered, the monolayer is still an indirect-band-gap semiconductor, but with a band gap of 0.99 eV. 

\begin{figure}[htbp]
\centering
\includegraphics[width=85 mm]{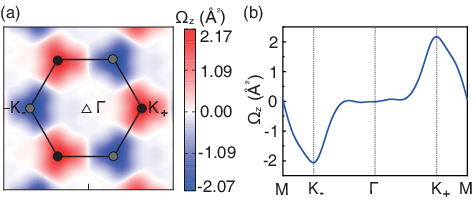} 
\caption{\label{FIG. 3}
Calculated Berry curvature of the H$^{\prime}$-Co$_2$CF$_2$ monolayer (a) over the two-dimensional reciprocal space and (b) along high symmetry lines. In (a), the first Brillouin zone is bounded by a hexagon.}
\end{figure}

Given that there are two inequivalent valleys with considerable valley splittings in the electronic band structure of the H$^{\prime}$-Co$_2$CF$_2$ monolayer, we further explore possible valley-contrasting physics. Fig.~\ref{FIG. 3} (a) and (b) show the Berry curvature of  all occupied bands over the two-dimensional Brillouin zone and along high symmetry lines, respectively. According to these figures, it is found that two peaks appear at $K_\pm$ valleys, with opposite signs. That is, the Berry curvature is indeed valley dependent due to the intrinsic inversion symmetry breaking in the H$^{\prime}$-Co$_2$CF$_2$ monolayer. Moreover, because of 
the valley degeneracy splittings, the extrema of the Berry curvature at two valleys has different absolute values. The valley-dependent Berry curvature will further lead to valley-spin-related quantum transports. 
Under the actions of an in-plane electric field and the carrier doping at two valleys, two opposite anomalous Hall currents will be driven by the Berry curvature with the same spin but from different valleys, demonstrating a single-spin version of the valley Hall effect. Besides, the valley splittings allow for the carrier doping at a certain valley, for example, the hole doping of the valence band at $K_-$ valley. That will give rise to a single anomalous Hall current, with certain indices of valley and spin.

\subsection{\label{sec:level2} Magnetic and electric controls of the valley splittings}
\begin{figure}[htbp]
\centering
\includegraphics[width=87 mm]{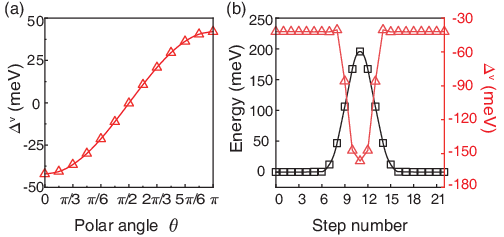} 
\caption{\label{FIG. 4}
Magnetic and electric modulations of the valley splitting of the valence band. (a) The change of the valley splitting with the magnetization direction. $\theta$ is the polar angle with respect to the $z$ axis. (b) The changes of the total energy and the valley splitting with the reversal of the electric polarization.}
\end{figure}

Considering that the H$^{\prime}$-Co$_2$CF$_2$ monolayer is found to be a ferrvalley material in the above section and it has ferromagnetic and ferroelectric orders as well, we will further study the couplings between these multiferroic orders. We first study the effect of the magnetization on the valley splittings. Fig.~\ref{FIG. 4} (a) shows the evolution of the valley splitting of the valence band as a function of the magnetization direction. With rotating the magnetization from the +$z$ direction ($\theta$=0) to the in-plane direction ($\theta$=$\pi$/2), the valley splitting is always negative, indicating that the valence band state at $K_+$ valley has a lower energy than that at $K_-$ valley. Meanwhile, the magnitude of the valley splitting gradually decreases from the maximum of 42 meV to zero. As we continue rotating the magnetization to -$z$ direction ($\theta$=$\pi$), the valley splitting changes the sign and gradually increases to +42 meV, where the positive sign means the valence band state at $K_+$ valley become higher than that at $K_-$ valley. Therefore, both the magnitude and the sign of the valley splitting can be tuned by rotating the magnetization direction. Moreover, the magnetizations along the $\pm z$ axis lead to the valley splittings with the same magnitude but opposite signs, which can be well understood from the view of the symmetry. The opposite magnetizations are related by the time-reversal operation, and the operation also relates $K_\pm$ valleys. The reversal of the magnetization gives rise to an exchange of the two valleys and correspondingly opposite valley splittings. 

We then study the effect of the electric polarization on the valley splittings. Fig.~\ref{FIG. 4} (b) shows the changes of the total energy and the valley splitting of the valence band during the reversal of the electric polarization. In the process of the ferroelectric switching, two degenerate H$^{\prime}$ phases are chosen as the initial and final states. They are related by the horizontal mirrior operation, and thus they have different vertical positions of the C atomic layer with respect to the middle of the H$^{\prime}$-Co$_2$CF$_2$ monolayer and  opposite out-of-plane electric polarizations. When considering the vertical shift of the C atomic layer as the possible pathway of the ferroelectric switching, the paraelectric H phase is chosen as an intermediate state. The calculated energy barrier between ferroelectric phase and paraelectric phase is 196 meV per formula unit, which agrees with the previous result \cite{19}. 

Meanwhile, along the pathway of the ferroelectric switching, the valley splitting is changed as well. The magnitude of the valley splitting increases from 42 meV at the H$^{\prime}$ phase to 157 meV at the H phase and then decreases back to 42 meV at the other H$^{\prime}$ phase. Furthermore, we demonstrate the band structures of a representative intermediate state and the H phase in the Supplemental Material (SM). The intermediate state exhibits a valley splitting of -86 meV, which falls between the values of the valley splittings of the H$^{\prime}$ and H phases. 
Since intermediate states can be realized experimentally by applied electric field for ferroelectric materials, the electric field is expected to modulate the atomic structures of the Co$_2$CF$_2$ monolayer and the magnitudes of the associated valley splittings. However, unfortunately, the sign of the valley splitting can't be tuned by the ferroelectric switching, unlike the magnetization reversal. This is because two degenerate H$^{\prime}$ phases are related by the horizontal mirror operation that can't exchange the valleys.

Above we demonstrate the changes of the valley splitting of the valence band with the magnetization rotation and ferroelectric switching. The changes of the valley splitting of conduction bands are also provided in SM. Similar to the case of the valence band, the valley splitting of conduction bands are tunable for both the magnitude and sign by rotating the magnetization, while the electric means can only modulate the magnitudes of the valley splittings.

\section{DISCUSSIONS} 
\begin{figure}[htbp]
\centering
\includegraphics[width=95 mm]{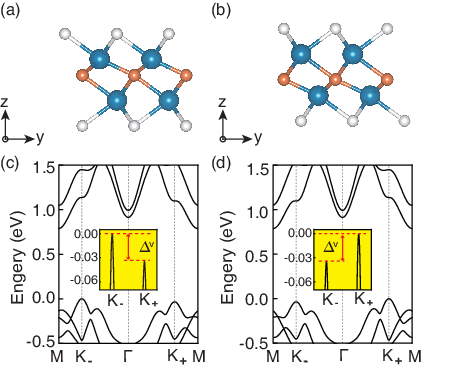} 
\caption{\label{FIG. 5}
Atomic structures and electronic band structures
of two degenerate T$^{\prime}$-Co$_2$CF$_2$ monolayers. (a) and (b) The side views of the atomic structures. (c) and (d) Corresponding electronic band structures. The insets zoom in the valence bands at $K_\pm$ valleys.}
\end{figure}

Although the sign change of the valley splitting can't be realized by the electric means in the above Co$_2$CF$_2$ monolayer, it is possible to occur in another phases of the Co$_2$CF$_2$ monolayer. In addition to the H and H$^{\prime}$ phase, the MXene monolayers also have T and T-like phases \cite{19,31}. We then discuss the ferroelectric switching and associated valley splittings of the Co$_2$CF$_2$ monolayer in the 
meta-stable T$^{\prime}$ phase. Fig.~\ref{FIG. 5} shows atomic structures and corresponding electronic band structures of two degenerate T$^{\prime}$ phases. As shown in Figs.~\ref{FIG. 5} (a) and (b), the middle Co-C-Co atomic layers of the T$^{\prime}$ phases have a similiar stacking order with those of the T phase and the T-MoS$_2$ monolayer, i.e. each C atom is bonded to six neighboring Co atoms forming a octahedron. Different from the MXenes in the T phase with a fcc-like stacking pattern for all atomic layers, the outmost F atomic layers of the T$^{\prime}$ phase exhibit in-plane shifts. Moreover, the T$^{\prime}$ phases have no inversion symmetry, which leads to a nonvanising electric polarization. The calculated electric polarization has a magnitude of 2.2 pC/m, which is in the same order of magnitude as two-dimensional sliding ferroelectric materials \cite{POLARIZATION1,POLARIZATION2,polarization3}. The electric polarizations of two degenerate T$^{\prime}$ phases are opposite, since they are related by the inversion symmetry. The energy barrier of the ferroelectric switching for two degenerate T$^{\prime}$ phases is also calculated, with a paraelectric T phase as an intermediate state, which is shown in SM.  In Figs.~\ref{FIG. 5} (c) and (d), the two degenerate phases also exhibit clear valley structures at $K_\pm$ for the valence band. They have opposite valley splittings, with values of -35 meV and 35 meV, respectively. This suggests that the sign of the valley splitting for the T$^{\prime}$ phase is tunable by the electric field driven ferroelectric switching. 

Although we focus on the Co$_2$CF$_2$ monolayer above, it is noted that there are a large number of two-dimensional materials. Among them, we can search for another candidates with possible couplings between ferrovalley, ferromagnetization and ferrelectricity, in order to realize the magnetic and electric controls of the valleys. For example, the ferroelectric Hf$_2$CF$_2$ monolayer \cite{31}, as another MXene material, also exhibits $K_\pm$ valleys in its electronic band structure (see SM). Though the valley splitting is absent in the monolayer due to its nonmagnetic ground state, the spin splittings of band edge states at $K_\pm$ valleys and associated spin-valley locking can be modulated by the reversal of the ferroelectric order. Moreover, not only the ferroelectric and ferromagnetic orders can be considered to be coupled with ferrovalley, the couplings between valley and antiferroelectric/antiferromagnetic orders are worth further studying as well.
\vspace{-3mm}

\section{CONCLUSION}
In summary, we have investigated multiferroic Co$_2$CF$_2$ monolayers in the H$^{\prime}$ and T$^{\prime}$ phases. Besides the ferroelectricity and ferromagnetism, they are also ferrovalley materials, with
considerable valley degeneracy splittings in their electronic band structures and with valley contrasting Berry curvature. Furthermore, it is found that the valley splittings can be modulated by the magnetization direction and applied electric field. Besides, the magnetic or electric controls of the valley-related properties can also apply to another MXene monolayers, such as Hf$_2$CF$_2$. Our study broadens the choices of valley materials and provides mutiple manipulations of the valley structures based on the couplings between ferroelectricity, ferromagnetism and ferrovalley. 

\begin{acknowledgments}
We are grateful to Erjun Kan, Chengxi Huang and Chenhan Liu for valuable discussions. We are supported by the National
Natural Science Foundation of China Grant 11904173
and the Jiangsu Specially-Appointed Professor Program.
\end{acknowledgments}

\nocite{*}

\providecommand{\noopsort}[1]{}\providecommand{\singleletter}[1]{#1}%

\end{document}